\documentclass[aps,twocolumn]{revtex4}

\usepackage{amsmath}  
\usepackage{amsfonts} 
\usepackage{graphicx, subfigure} 

\usepackage{color}
\definecolor{r}{rgb}{1,0,0}
\newcommand{\RS}{\color{black}}
\usepackage{epsfig}
\definecolor{g}{rgb}{0,1,0}

\definecolor{b}{rgb}{0,0,1}

\definecolor{m}{rgb}{1,0,1}

\begin{document}

\title{Marangoni flow in freely suspended liquid films}
\author{T. Trittel$^1$, K. Harth$^{1,2}$, 
Christoph Klopp$^1$, R. Stannarius$^1$
}
\affiliation{$^1$Institute of Experimental Physics, Otto von Guericke University, 39106 Magdeburg, Germany,
 $^{2}$Universiteit Twente, Physics of Fluids and Max Planck Center for Complex Fluid Dynamics, P.O. Box 217,
7500 AE Enschede, The Netherlands}

\date{\today}

\begin{abstract}

We demonstrate controlled material transport driven by temperature gradients in thin freely suspended smectic films.
The films with submicrometer thicknesses and lateral extensions of several millimeters were studied in microgravity during
suborbital rocket flights.
In-plane temperature gradients cause two specific Marangoni effects, directed flow and convection patterns.
At low gradients, practically thresholdless, flow transports material with a normal (negative) temperature coefficient of the
surface tension, $d\sigma/dT<0$, from the hot to the cold film edge. That material accumulates at the cold film border.
In materials with positive temperature coefficient, $d\sigma/dT>0$, the reverse transport from the cold to the hot edge is observed.
We present a model that describes the effect quantitatively.
\end{abstract}

\maketitle

{\RS Flow induced by capillary forces in thin fluid films has attracted scientific interest since the middle of 19$^{th}$ century \cite{Thomson1855}. It brings about not only the well-known B\'enard-Marangoni hexagonal convection patterns, but can also cause large-scale convection \cite{Vanhook1997,Bestehorn2003}. Marangoni flow plays a role in the evaporation dynamics of droplets
\cite{Deegan1997,Xu2007}, or bursting of bubbles \cite{Poulain2018}. One can exploit thermocapillary forces, e.~g., for a controlled manipulation of microfluidic systems and microdroplets \cite{Kotz2004,Farahi2004,Cordero2008,MacIntyre2018}.

In all these experiments, the fluid layers are in contact with a liquid pool or solid substrate, and surface forces create shear flow.
In contrast, freely suspended smectic films can be prepared without substrate, much like soap films. Such films can reach aspect
ratios (width:thickness) above $10^6$:1. Flow is restricted to the film plane, and no gradients exist normal to that plane.
Thus, thermocapillary forces can be much more effective than in substrate-supported films.
We demonstrate thermally driven macroscopic material transport in such quasi two-dimensional (2D) fluids.}

\begin{figure}[htbp]
\includegraphics[width=0.8\columnwidth]{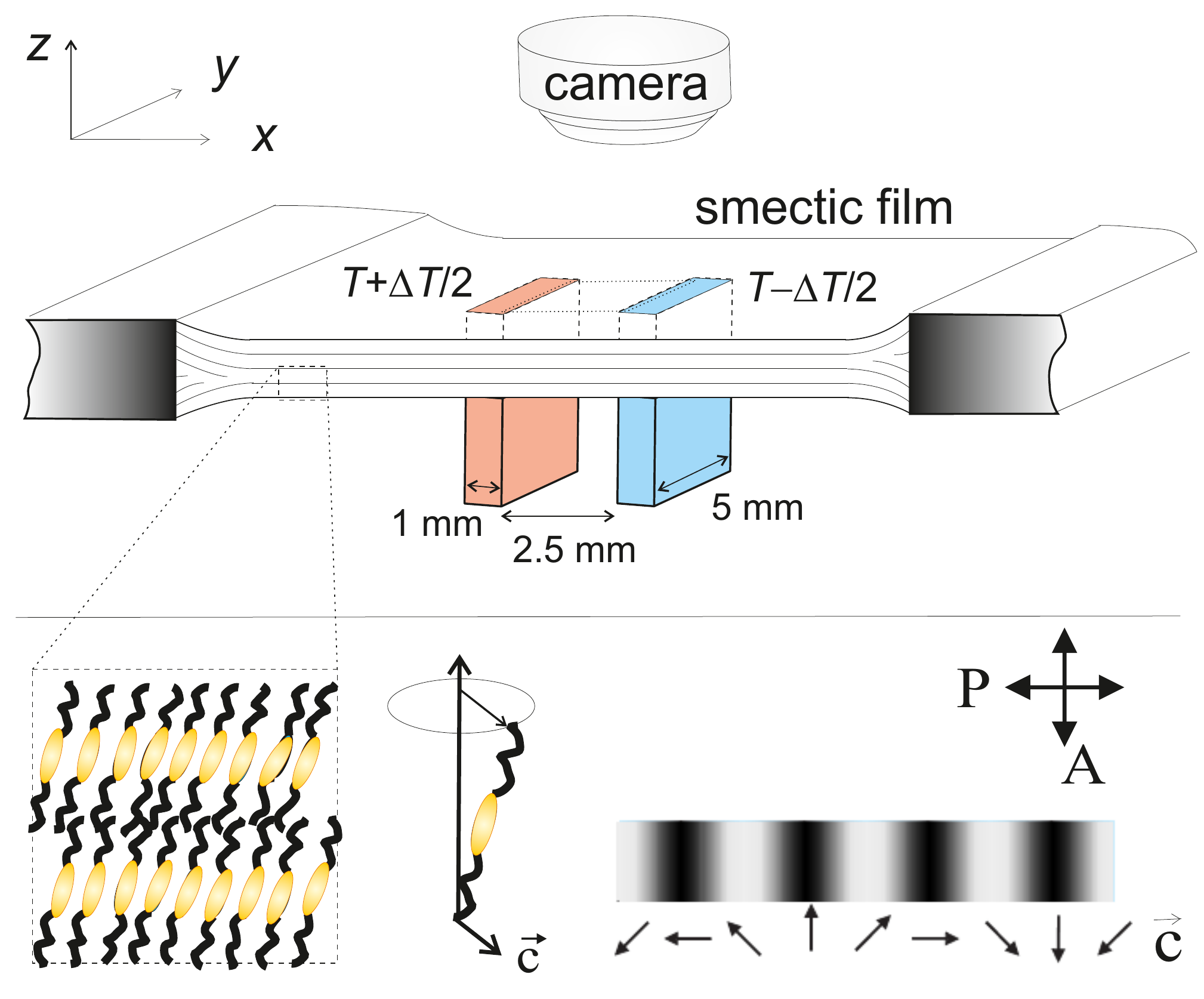}
  \caption{Color online: The experimental geometry is sketched in the top part. Front and rear edges of the frame are omitted.
  The total film area is 13~mm $\times$ 10~mm, the field of view is 6~mm $\times$ 5~mm. Two pads with rectangular cross sections, in contact with the film,
  separated by $d=2.5$~mm, generate linear temperature gradients. These pads are
  set to temperatures $T_0 \pm\Delta T/2$, the surrounding film holder is kept at $T_0$.
  The bottom drawing depicts the molecular structure of the SmC film, and the
  definition of the c-director. The bar sketches the optical reflectivity under crossed polarizers (P,A)
  for different c-director orientations.
  }\label{fig:setup}
\end{figure}

Figure~\ref{fig:setup} sketches one of the simplest mesophases, smectic C (SmC), in the geometry
of a freely suspended film. 
A remarkable amount of literature describes hydro\-dynamics and director field structures of such films
(e.~g. \cite{Oswald2005,Cladis1985,Cladis1993,Mutabazi1992,Link2000,Stannarius2006,Zoom2010,Eremin2011}),
pattern formation \cite{Maclennan1990,Morris1990,Langer2000}, inclusions in the films (see Refs. in \cite{Bohley2008}), shape dynamics
(e.~g.~\cite{Benamar1998,Muller2006,May2012}), rupture \cite{Muller2007,Trittel2013,Trittel2017} and other aspects.
Commonly, flow fields are described by a 2D Stokes equation for incompressible fluids, neglecting inertia.
Almost all experiments so far were performed under isothermal conditions, only few studies
reported effects of thermal gradients in the film plane \cite{Godfrey1996,Birnstock2001}.
Thermally driven motion in such films remains an challenging and so far unsolved problem.

In horizontal films, one can neglect gravity effects in flow processes.
However, this is justified only when the film and the setup are isothermal, or when the setup is evacuated.
With thermal gradients, air convection is practically unavoidable. Air drag induces flow \cite{Birnstock2001},
even in horizontal films. It can be inhibited by evacuation of the setup.
Godfrey and van Winkle \cite{Godfrey1996} investigated films with thermal gradients in vacuum. 
In that case, however, the film quickly adopts the uniform ambient temperature. Radiation loss of the film is much stronger
than thermal conduction. Even if film edges are at different temperatures, thermal gradients are limited to narrow
meniscus regions. Convection was found already at temperature differences $\Delta T$ below 0.1~K
across a 3.1~mm film. At 0.32~K/mm, 
flow velocities of 35~$\mu$m/s were measured \cite{Godfrey1996}. However, this convection was driven by the
strong thermal gradients in the menisci, and not related to gradients in the film.

For the study of the genuine effects of thermal gradients in the films, one needs to keep the films in contact with
ambient air. The thermal diffusivity of air $\alpha_{\rm air} \approx 22$ mm$^2$/s is large compared to $\alpha_{\rm LC} \approx 0.06$
mm$^2$/s typical for a SmA liquid crystal \cite{Marinelli1996}. It rapidly establishes a uniform temperature gradient in the region between the thermocontacts (Fig.~\ref{fig:setup}). For a  $2.5$~mm gap, the typical time is about 300 ms. With ambient air, however, micro\-gravity ($\mu g$) is
needed to suppress buoyancy driven convection.
In $\mu$g, all thermally driven motion can be attributed to Marangoni effects, arising from the temperature dependence of the smectic surface tension $\sigma(T)$.

\begin{figure}[htbp]
	\includegraphics[width=0.95\columnwidth]{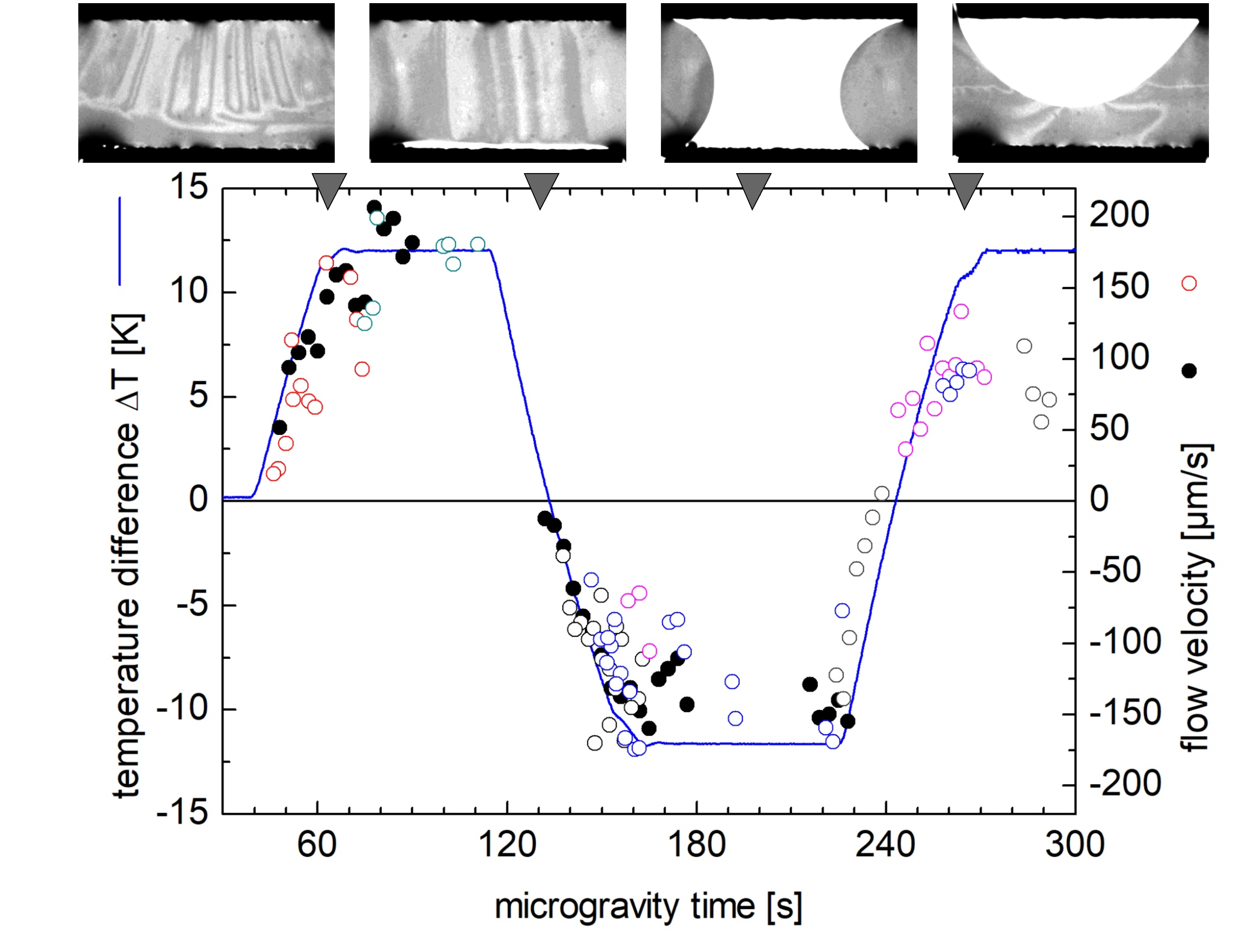}
	\caption{Temperature protocol during the first 300~s of microgravity (TEXUS 55), and the velocity component $v_x$ determined
    from the texture displacement. Different colors represent different positions in the film. The accuracy is $\pm 20~\mu$m/s.
}
	\label{fig:protocol}
\end{figure}

The experiments were performed on suborbital rocket flights at Esrange (Sweden) with TEXUS 52 on April 27, 2015, and TEXUS 55 on May 18, 2018.
Each flight provided approximately 360~s of $\mu g$.
The LC material was commercial 5-n-Octyl-2-(4-n-octyloxyphenyl) pyrimidine ({\em SYNTHON Chemicals}),
referred to as 8PP8. Its mesomorphism is
isotropic 69$^\circ$C nematic 62$^\circ$C SmA 55.5$^\circ$C SmC 28.5$^\circ$C cryst.
The SmC phase can be supercooled below room temperature.
The SmC film was drawn during the first 30 sec of the $\mu$g phase. It had a final area of 10~mm~$\times$~13~mm.
A homogeneous film thickness $h$ was established within a few seconds, $h=(535 \pm 10)$~nm (TEXUS 52) and $h=(170 \pm 20)$~nm (TEXUS 55)
were determined interferometrically.
The temperature gradient between the pads was varied during the microgravity period (TEXUS 55) as shown in Fig.~\ref{fig:protocol}.
Temperatures were controlled by Peltier elements, they could be ramped at maximum rates of 0.3~K/sec.
$T_0=55 ^\circ$C was chosen in the SmC range.
The thermopads were short-circuited by a thin bond wire to avoid electrostatic effects.
We observed the film region between the thermopads with a video camera (resolution $5~\mu$m/pixel) in polarized reflected light.
Instead of tracer motion, the drift of Schlieren textures of the non-uniform c-director (optic axis) in the film was exploited to
extract the velocity fields.
Here, we report the evaluation of the TEXUS 55 experiment, the data of TEXUS 52 were consistent and in reasonable agreement with the
results described below.

\begin{figure}[htbp!]
	\centering
	\includegraphics[width=0.85\columnwidth]{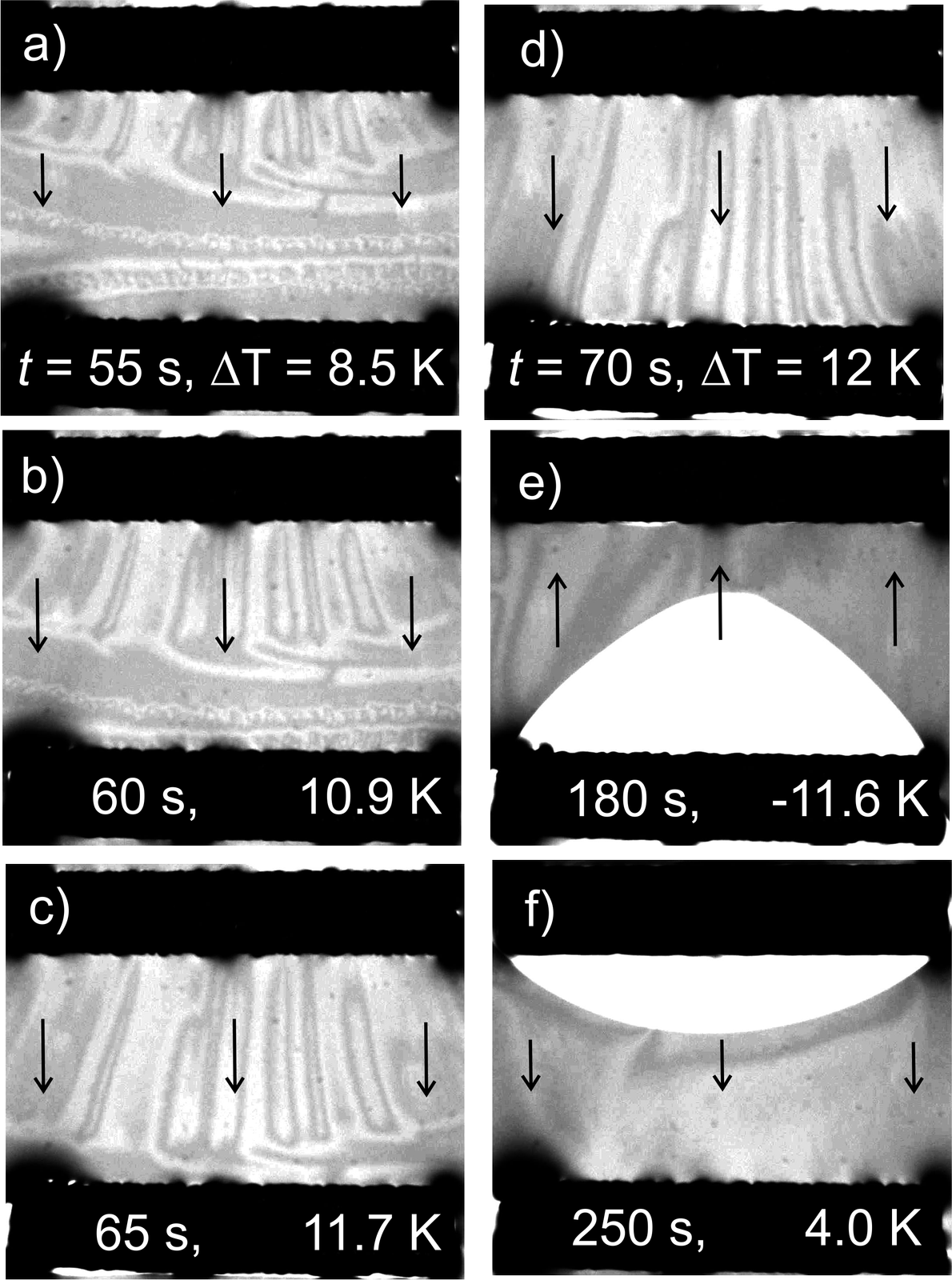}
	\caption{8PP8 textures (TEXUS 55) in the region between the pads: a-d) Texture transported by flow from the hot pad (top) to the cold pad (bottom), e) reverse  flow after reversal of $\Delta T$, thicker film is transported away from the former cold plate (bottom), f) film after second reversal of $\Delta T$ and flow direction.
}
	\label{fig:textures}
\end{figure}

Figure \ref{fig:textures} shows selected views of the film region between the pads (black bars) at gradients up to $\pm 4.8$ K/mm.
The textures evidence a uniform flow of the film from the hot to the cold plate.
The film thickness thereby remains constant, the transported material accumulates at the cold pad and forms a thick
wedge-shaped zone. The measured speed is shown in Fig.~\ref{fig:protocol}. Within the first minute, approximately six times the film
volume between the pads is transported to the cold pad.
Reversal of the temperature gradient reverses the flow (Fig.~\ref{fig:textures} e).
Now, a thicker but uniform film region (island) is carried with the flow from the now hot to the now cold pad.
The flow speed is lower because more film material is carried, and the flow profile is more complex because of the additional line
tension of the island boundary. The thicker island finally reaches the cold pad. Another reversal of the temperature gradient again
reverses the flow direction, the island drifts back (Fig.~\ref{fig:textures} e). In this experiment, directed material transport
across the film is achieved by thermal gradients. The flow velocity follows the temperature gradient ramp, without measurable delay.
Convection rolls set in only at larger temperature gradients, as applied in the TEXUS 52 experiment.

The effect can be quantitatively explained as follows: A constant flow of the film replaces local cold film material by warmer one,
thereby the temperature profile remains roughly linear, but is shifted towards the cold edge. This reduces the surface energy.
The speed is determined by the energy needed to produce layer dislocations at the cold film edge where the smectic material is collected,
and energy needed to remove dislocations at the hot edge.
In each film segment, there is an equilibrium of heat flow in the film by material drift and conduction, and heat transport
in the surrounding air which involves a small temperature gradient normal to the film. We will estimate these contributions quantitatively
to show that the model is reasonable and that typical parameters for our material are suited to describe the experiment correctly.
For simplicity, we focus on the data of the first heating cycle, with the uniformly thick film.

An estimate of temperature profiles and the order of magnitude of the related heat transport shall demonstrate that the assumptions are
quantitatively reasonable. We use a film thickness of $h=170$ nm and a temperature difference
$\Delta T=12$~K across the $d=2.5$~mm gap. It produces a profile $T(x)=T_0+\theta (x-d/2)$ with
$\theta= dT/dx=-4800$~K/m. The pad width $b=5$~mm roughly defines the width of the flowing region.
The flow velocity $v_x\approx 150~\mu$m/s is taken from Fig.~\ref{fig:protocol}.
The heat flow $P$, transported by thermal diffusion in the film with an assumed heat conductivity
 $\lambda=0.13$ Wm$^{-1}$K$^{-1}$ \cite{Marinelli1996} is negligible,
$
P = \lambda b h \theta \approx  0.5~\mu$W. 
Air layers of, say, $h'=1$~mm thickness on both sides conduct 
$dP'=2 b h' \lambda' \theta \approx 1.2$~mW along $x$ ($\lambda'\approx 0.028$~Wm$^{-1}$K$^{-1}$).
The heat transported by drift $v_x$ of the film with an approximate heat capacity of $c\approx 2  \cdot 10^{6}$ Jm$^{-3}$K$^{-1}$
is roughly $ P_D = b h c v_x T\approx 80~\mu$W.
Air in adjacent layers transports about $P'_D = 2 b h' c' (v_x/2) T \approx 270~\mu$W, where we assume for simplicity a
linear profile $v_x(z)$ over the height $h'$. Since this is an order of magnitude estimation, a factor of 2-3 in $h'$ is not
relevant.

In a linear gradient, without flow, the heat transported by diffusion along $x$ into and out of a vertical slice is
balanced.
However, the flowing layers inject excess heat power $d\,(P_D+P'_D) =(  h c  +  h'c' )bv_x\theta \cdot  {dx}$ into each
slice $dx$. In a stationary state, this heat is dissipated into the surrounding air by a gradient vertical to the film,
$\Delta T_\perp/h' =(T^{\ast}(x) -T(x))/h'$, where $T^\ast(x)$ is the elevated film temperature compared
to the state without flow.

With $d(P_D+P_D') = 2\lambda' b \Delta T_\perp/h' dx$, the required gradient is
\begin{equation}
\frac{\Delta T_\perp}{h'} =\frac{(hc+h'c')v_x \theta}{2\lambda'} \approx 20~{\rm K/m}.
\label{deltaT}
\end{equation}
Thus, a global elevation of the film temperature of the order of $\Delta T_\perp\approx 20$~mK relative to the non-flowing film is sufficient to reach a stationary profile. This lowers the specific surface energy by $\Delta E_{\rm surf} = 2 d\sigma/dT\cdot \Delta T_\perp$.
With $d\sigma/dT \equiv \Sigma \approx -7.9\cdot 10^{-5}$~N/(m$\cdot$K), the reduction amounts to
$ \approx -3.2~\mu$N/m. 

\begin{figure}[htbp]
	\centering
	\includegraphics[scale=0.6]{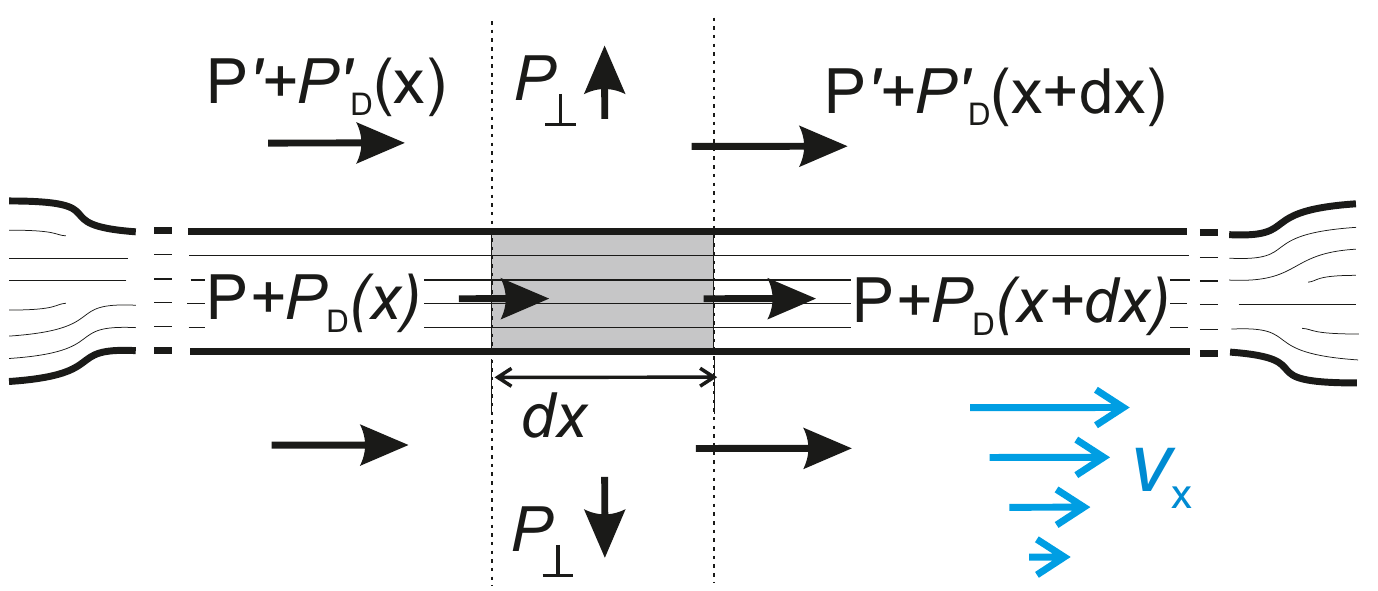}
	\caption{Flow profile and heat flow in and around the freely suspended film.
    At the sides, the meniscus is sketched with dislocations that are related to the film thickness gradients.}
	\label{temperature_flow}
\end{figure}

The kinetic energy per unit area of the flowing film and air layers is of the order of
$\Delta E_{\rm kin}= \left( \rho h/2 +\rho'h'/3\right) v_x^2\approx  10$ pN/m (in air, the average squared velocity
is $v_x^2/3$ on each film side). The gain of surface energy is
orders of magnitude higher than the kinetic energy needed to produce flow.
The relation between $\Delta E_{\rm surf}$ and $v_x$ is
\begin{equation}
\Delta E_{\rm surf} = -2\frac{d\sigma}{dT}\Delta T_\perp = -\frac{d\sigma}{dT}\, h'\frac{ (h c+h'c')  \theta}{\lambda}\; v_x
\end{equation}
This means that the onset of the transport is practically thresholdless. 
Any acceleration of $v_x$ leads to a larger shift of the temperature gradient and thus to a further reduction of surface energy.
Thus the film would continuously accelerate if there was no counteracting dissipative mechanism.
Except at the lateral sides of the flowing area, there is no shear flow involved.
This is clearly seen in Fig. \ref{fig:textures}b,c, where a nearly straight front passes the film.
Unlike convective rolls, the uniform flow between the pads does not dissipate energy.
Some dissipation occurs in the adjacent air, but the power dissipated per area
$P_{\rm shear}\approx 2 \eta' v_x^2/h'$ is only of the order of $10^{-10}$~W/m$^2$,
and is thus negligible in our estimation.
The total gain from surface energy reduction in the area between the pads is
$ 2 |\Sigma| \Delta T_\perp b d \approx 40$~pJ in our experiment.

The dissipation that limits the flow speed occurs almost exclusively at the film edges.
It is generated by the process that stacks the film material at the cold edge, and the process which removes smectic material
from the meniscus at the warm edge.
Thereby, layer dislocations are created on one side and destroyed at the other side of the film (Fig.~\ref{temperature_flow}).
A dynamic equilibrium is formed, where the global gain in surface energy and the dissipation of energy in the menisci are balanced.
It is possible to estimate this balance quantitatively.

For that purpose, we calculate the forces created by the surfaces at both edges. The contacts are at temperatures
$T_1=T_0+\Delta T/2$ and $T_2=T_0-\Delta T/2$, respectively. The difference between the corresponding surface tensions
is $\Delta \sigma =\Sigma \Delta T \approx 9.5\cdot 10^{-4}$~N/m. This provides a force
 $2\Delta \sigma$ per meniscus length (two film surfaces, on top and bottom) acting on the film, equivalent to a pressure difference
 $\Delta p=2\Delta \sigma/h  \approx 11 $~kPa between the film cross sections at the contacts.
We compare this to the friction force of the film moving in air. In the film region between the pads, there are practically no shear gradients, we assume a uniform transport velocity $v_x$ (see experiment).
Supposing, as above, roughly linear velocity gradients in the air layers of $v_x/h'$, the air friction force per film width amounts to
$ 2d\eta' {v_x}/{h'} 
\approx 1.5\cdot 10^{-9}{\rm N/m}$
(air viscosity $\eta'\approx 2\cdot 10^{-5}$~Pas), with a negligible contribution to the pressure at the film edges,
of the order of 0.1~Pa.

The mechanism damping the flow must be sought in the forces the smectic material in the meniscus develops to counteract inflow and outflow (creating, moving and removing dislocations).
The necessary pressure to keep a stationary flow rate must be of the order of a few kPa.
Oswald und Pieranski \cite{Oswald2005} derived an equation for the related dissipation (energy loss per time and meniscus length)
\begin{equation}
\label{eq_dissip}
\phi_m= h v_x^2 \frac{1}{m},
\end{equation}
where $m$ is a quantity characterizing the mobility of dislocations, its unit is the inverse of a viscosity per length.
On the basis of measurements with the smectic material 4n-octyloxy-4-cyanobiphenyle (8CB) at 28$^\circ$C, the authors
reported a value of $m= 4.44\cdot 10^{-7} {\rm cm}^2 {\rm s/g}\equiv 4.44\cdot 10^{-8}$~m/(Pa\,s) \cite{Picano2000}.
The pressure with which the meniscus opposes an accretion or extraction of smectic material is \cite{Oswald2005}
\begin{equation}
p_m=\frac{v_x}{m}.
\end{equation}

For our experiment, this means that $\Delta p=2p_m$, the factor 2 accounts for inflow and outflow at the hot and cold edges.
On the basis of the measured flow velocity $v_x$, we obtain
\begin{equation}
m=v_x h/\Delta\sigma = 2.7\cdot 10^{-8} ~{\rm m/(Pa\,s)}
\end{equation}
within a 30~\% uncertainty range.
This dislocation mobility is of the same order of magnitude as the 8CB value reported by Picano et al. \cite{Picano2000}.
Taking into account that Picano et al. used a different material, that their temperature was 20 K lower than in
our experiment, and that the material parameters $c,\lambda$ were estimated from literature data for similar mesogens, this is a
very reasonable result.
The data obtained at the TEXUS 52 mission are qualitatively comparable but differ quantitatively. The
film was much thicker there (585 nm) and the flow velocity was substantially smaller ($\approx 25~\mu$m/s at $\Delta T = 15$~K).
A lower velocity of thicker
films is consistent with our model. The mobility estimated from the TEXUS 52 data was only $1.6\cdot 10^{-8} ~{\rm m/(Pa\,s)}$.
Because of the more complex temperature protocol in that experiment, which was necessary for the determination and selection of
the relevant parameter ranges, that experiment was quantitatively less accurate.

\begin{figure}[htbp]
	\centering
	\includegraphics[width=0.85\columnwidth]{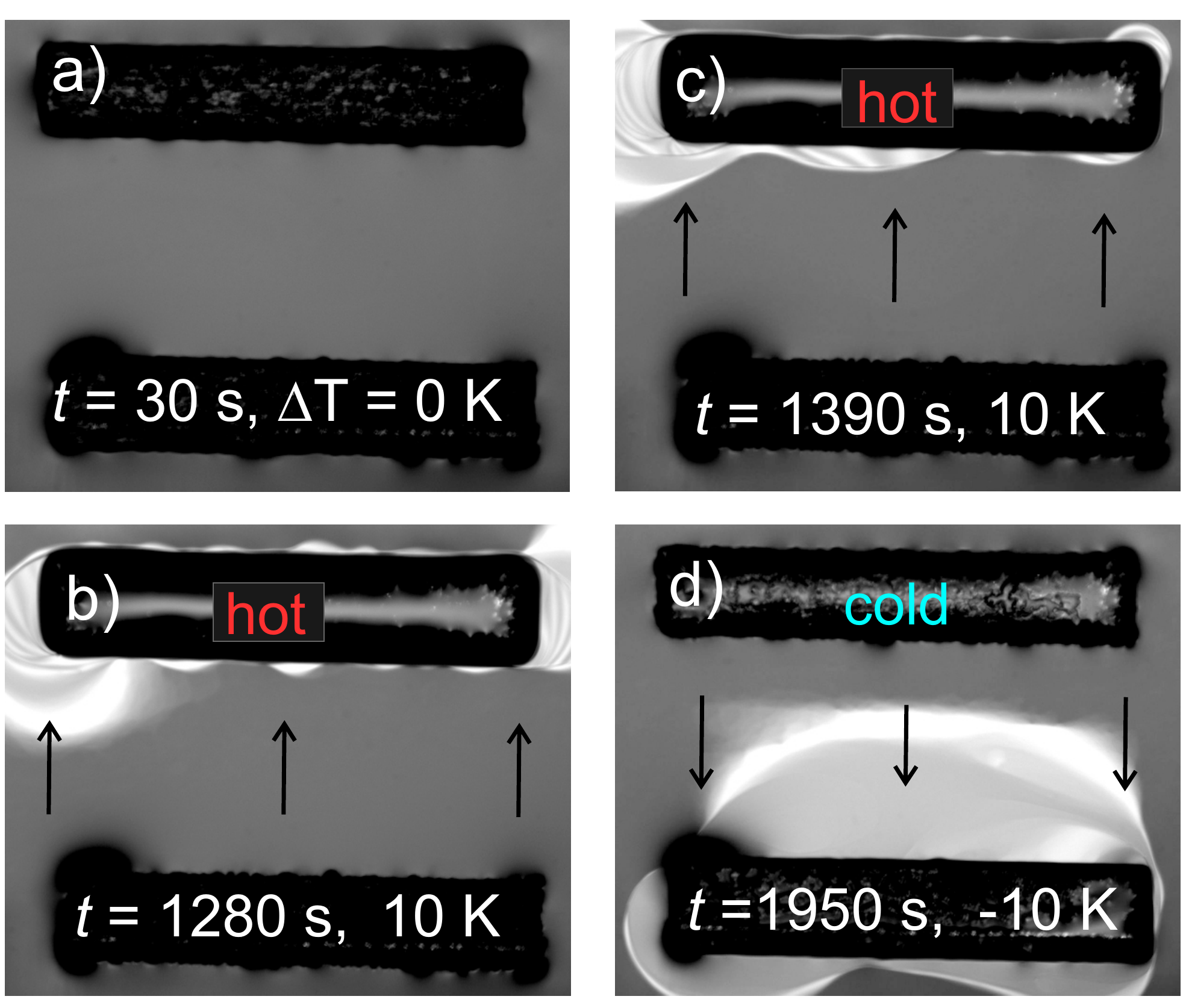}
	\caption{5O.6 film in the region between the pads: a) initial uniform film, b,c) accumulation of thick film at the hot pad (top)
 and d) after reversal of the temperature gradient. The material is in the smectic A phase, thus no textures are visible.
 Bright regions represent thicker film, interference fringes in (b,c) evidence thickness gradients.
 Because air convection could not be completely avoided, the accumulated thicker region at the hot pad tends to wobble (b,c).}
 \label{fig:5O.6}
\end{figure}

The litmus test of our model was an experiment with the same setup and the mesogen
N-(4-n-Pentyl\-oxy\-benzylidene)-4'-Hexylaniline (5O.6) in the ground lab.
This material has an unusual positive temperature coefficient $d\sigma/dT\approx +7.7\cdot 10^{-5}$~N/(m$\cdot$K) \cite{Schuring2001}. Consequently,
one may expect
that the surface energy is lower at the cold edge and the temperature gradient is shifted by flow towards the hot edge. The lack of
availability of another suborbital rocket flight was not problematic in this qualitative experiment. Any buoyancy driven
air convection under normal gravity will lead to an upstream of air at the hot edge, and flow beneath the film to the cold edge,
and a downstream of the convection roll there. If such flow is present, it will tend to push film material towards the cold edge as in
the above described microgravity experiment. Actually, however, we clearly observed flow in the opposite direction (Fig.~\ref{fig:5O.6}).
Film material is transported away from the cold pad and it accumulates at the hot edge, Marangoni transport clearly dominates.
This observation does not only support our model, it also demonstrates that by proper selection of materials, one can control
a directed flow in either directions, with and against the temperature gradient, in freely suspended fluid films.
The results will help to interpret experiments performed on the ISS within the OASIS
project, where the motion of islands of smectic material on uniform background films (smectic bubbles) has been observed in
bubbles exposed to thermal gradients \cite{Clark2017}.

  Summarizing, our observations underline the necessity to account for Marangoni flows in all situations where smectic freely suspended films
are not under isothermal conditions. We presented a quantitative model that predicts the expected flow velocities and may provide the basis for potential microfluidic applications.

\section*{Acknowledgments}

The OASIS-Co project is supported by the German Aerospace Center (DLR) with grants 50WM1430 and 50WM1744, in the context of the project OASIS. The authors are particularly indebted to DLR for generous support and for making the TEXUS experiments possible, and to
AIRBUS DS for the construction and testing of the  equipment as well as their technical support during the TEXUS-52 and TEXUS-55 campaigns at Esrange. OASIS was funded by NASA Grant~NNX-13AQ81G. Alexey Eremin, Noel A. Clark, Joseph E. Maclennan, Cheol S. Park are cordially acknowledged for their decisive contributions to the OASIS microgravity experiments and for numerous stimulating discussions.

\bibliographystyle{unsrt}

\begin{thebibliography}{99}
\bibitem{Astrium}
The experimental setup was built and tested by AIRBUS DS (Astrium).
\end{thebibliography}

\end{document}